\documentclass[submission,Phys]{SciPost}
\usepackage{color,amsmath,amssymb,multirow,hyperref,booktabs,graphicx,mathtools} 

\graphicspath{{./figs/}}

\renewenvironment{thebibliography}[1]{
  \begin{oldthebibliography}{#1}
    \setlength{\itemsep}{0em}
    \setlength{\parskip}{0em}
}
{
  \end{oldthebibliography}
}

\DeclareMathOperator{\tr}{Tr}
\newcommand\one{\leavevmode\hbox{\small1\normalsize\kern-.33em1}}

\newcommand{\lag}{\mathcal{L}}

\newcommand{\ope}{\mathcal{O}}
\newcommand{\qqquad}{\qquad \qquad}

\newcommand{\gev}{\text{GeV}}
\newcommand{\tev}{\text{TeV}}

\newcommand{\iab}{\text{ab}^{-1}}

\def\slashchar#1{\setbox0=\hbox{$#1$}           
   \dimen0=\wd0                                 
   \setbox1=\hbox{/} \dimen1=\wd1               
   \ifdim\dimen0>\dimen1                        
      \rlap{\hbox to \dimen0{\hfil/\hfil}}      
      #1                                        
   \else                                        
      \rlap{\hbox to \dimen1{\hfil$#1$\hfil}}   
      /                                         
   \fi}

\newcommand{\be}{\begin{eqnarray*}}
\newcommand{\ee}{\end{eqnarray*}}

\newcommand{\bee}{\begin{eqnarray}}
\newcommand{\eee}{\end{eqnarray}}
\newcommand{\beeq}{\begin{equation}}
\newcommand{\eeeq}{\end{equation}}

\begin{document}

\begin{center}{\Large \textbf{
The Global Higgs Picture at 27 TeV
}}\end{center}
\begin{center}
Anke Biek\"otter\textsuperscript{1},
Dorival Gon\c{c}alves\textsuperscript{2},
Tilman Plehn\textsuperscript{1},
Michihisa Takeuchi\textsuperscript{3}, and
Dirk Zerwas\textsuperscript{4}
\end{center}

\begin{center}
{\bf 1} Institut f\"ur Theoretische Physik, Universit\"at Heidelberg, Germany\\
{\bf 2} PITT PACC, Department of Physics and Astronomy, University of Pittsburgh, USA \\
{\bf 3} Kavli IPMU (WPI), UTIAS, University of Tokyo, Kashiwa, Japan \\
{\bf 4} LAL, IN2P3/CNRS/ Orsay, France \\
biekoetter@thphys.uni-heidelberg.de
\end{center}

\begin{center}
\today
\end{center}


\section*{Abstract}
{\bf We estimate the reach of global Higgs analyses at a 27 TeV hadron
  collider in terms of Higgs couplings and in terms of a
  gauge-invariant effective Lagrangian, including invisible Higgs
  decays and the Higgs self-coupling. The new collider will indirectly
  probe new physics in the TeV range and allow for a meaningful test
  of the Higgs self-coupling also embedded in a global analysis.}

\vspace{10pt}
\noindent\rule{\textwidth}{1pt}
\tableofcontents\thispagestyle{fancy}
\noindent\rule{\textwidth}{1pt}
\vspace{10pt}

\newpage
\section{Introduction}
\label{sec:intro}

In record time Higgs physics has moved from a spectacular discovery of
a new particle to a systematic and comprehensive study of its
properties~\cite{Dawson:2018dcd}. The general development of hadron
collider physics into precision physics has been fueled by the
understanding of the LHC detectors, the control of perturbative QCD,
the ability to precisely simulate even complex LHC processes, and the
theoretical understanding of a perturbative electroweak Lagrangian as an
interpretation framework. This naturally leads to the question what
kind of precision we can reach with a 27~TeV hadron collider with an
attobarn-level integrated luminosity.
 
If we assume that the observed scalar is really Higgs-like,
specifically that it is responsible for electroweak symmetry breaking
and hence forms a doublet with the weak Goldstone modes, we can
interpret the LHC results in terms of an effective Lagrangian with
linearly realized electroweak symmetry
breaking~\cite{Leung:1984ni,Buchmuller:1985jz,DeRujula:1991ufe,Hagiwara:1993qt,Hagiwara:1993ck,Hagiwara:1995vp,GonzalezGarcia:1999fq,Grzadkowski:2010es,Passarino:2012cb,Brivio:2017vri}.
This defines the framework of some of the most interesting global
Higgs analyses based on
Run~I~\cite{Corbett:2015ksa,Corbett:2012ja,Banerjee:2013apa,Ellis:2014jta}
and even Run~II~\cite{Ellis:2018gqa} data. Because this effective
Lagrangian firmly links the Higgs and electroweak sectors, the global
analysis has to incorporate anomalous triple gauge boson measurements
from LEP~\cite{Brivio:2013pma,Falkowski:2015jaa} and the
LHC~\cite{Corbett:2015ksa,Falkowski:2015jaa,Falkowski:2016cxu,Berthier:2016tkq,Liu:2018pkg,Baglio:2017bfe,Franceschini:2017xkh}.
At the 13~TeV LHC the Higgs self-coupling can be neglected at the
typical precision of a global Higgs analysis~\cite{Kling:2016lay}. In
contrast, a 27~TeV hadron collider is expected to contribute a
meaningful measurement of Higgs pair
production~\cite{Goncalves:2018yva,Kilian:2018bhs,Bizon:2018syu,Homiller:2018dgu},
for example testing a possible first-order electroweak
phase transition as an ingredient to
baryogenesis~\cite{Grojean:2004xa,Reichert:2017puo}. Finally, we
include invisible Higgs decays in terms of an invisible branching
ratio following Ref.~\cite{Biekotter:2017gyu}.

For our brief study we start from the established Run~I limits of the
LHC~\cite{Corbett:2015ksa,Butter:2016cvz} and extrapolate them to an
upgraded LHC setup\footnote{Many aspects of our 27~TeV study are
  described in detail in these 8~TeV legacy
  papers~\cite{Corbett:2015ksa,Butter:2016cvz}, including a validation
  of the 8~TeV results. The \textsc{SFitter} error treatment is
  discussed in Ref.~\cite{Lafaye:2009vr}.}.  This provides a reliable
benchmark for such an energy upgrade with a large integrated
luminosity. First, we discuss the 27~TeV projections in terms of Higgs
coupling modifiers, motivated by an effective Lagrangian with a
non-linear realization of electroweak symmetry breaking, in
Sec.~\ref{sec:higgs}. In Sec.~\ref{sec:gauge} we then use a linear
realization to combine the global Higgs analysis with di-boson
data. In Sec.~\ref{sec:self} we discuss the benefits from including a
differential measurement of Higgs pair production in detail.

Eventually, this analysis should be combined with anomalous gauge
couplings to fermions, as well as electroweak precision
data~\cite{deBlas:2016ojx,Baglio:2017bfe,Franceschini:2017xkh,Ellis:2018gqa,Alves:2018nof,Biekotter:2018rhp}.
We leave such a more detailed analysis also including a larger set of
kinematic distributions~\cite{Englert:2015hrx} and an appropriate
validation on ATLAS and CMS results for the future.  For now, this
extrapolation based on our established and validated 8~TeV analysis
will provide a reliable first estimate. Such a conservative treatment
is in order, because dependent on the interpretation framework the
global Higgs analyses at a 27~TeV collider will rapidly enter
systematics-limited and theory-limited territory.

\section{Global Higgs analysis} 
\label{sec:higgs}

\begin{figure}[t]
\centering
\includegraphics[width=0.80\textwidth]{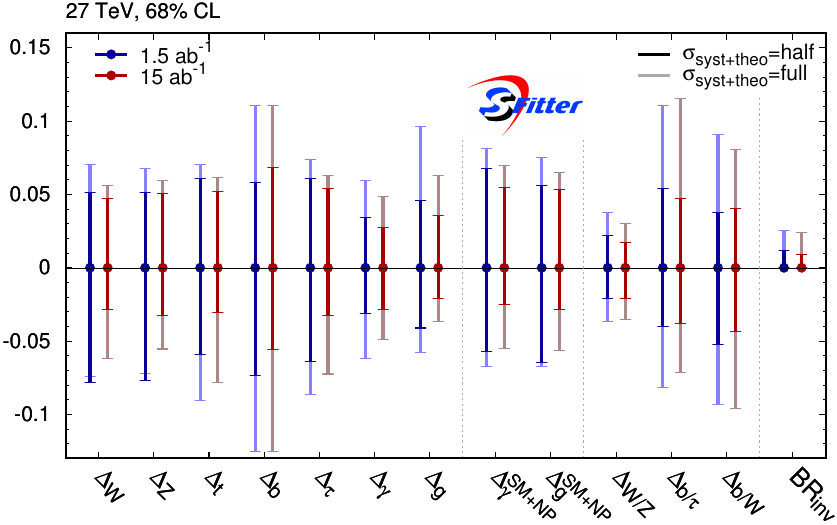}
\caption{Result from the global Higgs analysis in terms of coupling
  modifiers or non-linearly realized electroweak symmetry
  breaking. All limits are shown as profiled over all other
  couplings.}
\label{fig:delta}
\end{figure}

\begin{table}[b!]
\centering
\begin{tabular}{rr|rr}
\toprule
production  & $[\%]$ & decay & $[\%]$\\
\midrule
GF          & $10.2$ & $WW$ & $2.63$ \\
$qqH$          & $3.0$   & $ZZ$   & $2.63$ \\
$WH$           & $3.2$   & $\gamma\gamma$ & $3.31$\\
$ZH$           & $5.7$   & $b\bar{b}$ & $2.17$\\
$t\bar{t}H$    & $12.8$  & $Z\gamma$ & $7.33$\\
$HH$           & $18.$      & $\tau \tau$  & $2.78$\\
\bottomrule
\end{tabular}
\caption{Relative theory uncertainties for the different production
  and decay channels contributing to the global analysis. The numbers
  correspond to those quoted in Ref.\cite{deFlorian:2016spz}. 
  }
\label{tab:uncert}
\end{table}

Historically~\cite{Zeppenfeld:2000td,Duhrssen:2004cv,Lafaye:2009vr},
new physics effects on the SM-like Higgs couplings have been
parameterized as coupling modifiers
\begin{alignat}{6}
 g_x &= g_x^\text{SM} \; (1 + \Delta_x) \notag \\
 g_{g,\gamma} &
= g_{g,\gamma}^\text{SM} \; (1 + \Delta_{g,\gamma}^\text{SM} + \Delta_{g,\gamma} ) 
\equiv g_{g,\gamma}^\text{SM} \; (1 + \Delta_{g,\gamma}^\text{SM+NP} ) \; ,
\label{eq:def_delta}
\end{alignat} 
where the $\Delta_x$ can be directly translated into the
experimentally used $\kappa$ notation 
\begin{alignat}{6}
 \kappa_x &=  (1 + \Delta_x) 
\label{eq:def_kappa}
\end{alignat} 
at least modulo the treatment of
the tree-level couplings contributing to the loop-induced operators.
The corresponding global Higgs analysis is the main reason why we can
now claim that the observed Higgs boson closely follows the Standard
Model predictions.  In terms of a Lagrangian we can write this
hypothesis as~\cite{Corbett:2015ksa}
\begin{alignat}{5}
\lag 
= \lag_\text{SM} 
&+ \Delta_W \; g m_W h \; W^\mu W_\mu
+ \Delta_Z \; \frac{g}{2 c_W} m_Z h \; Z^\mu Z_\mu
- \sum_{\tau,b,t} \Delta_f \; 
\frac{m_f}{v} h \left( \bar{f}_R f_L + \text{h.c.} \right) \notag \\
&+  \Delta_g F_G \; \frac{h}{v} \; G_{\mu\nu}G^{\mu\nu}
+  \Delta_\gamma F_A \; \frac{h}{v} \; A_{\mu\nu}A^{\mu\nu} 
+ \text{invisible decays} \; .
\label{eq:lag_delta}
\end{alignat} 
This Lagrangian shifts all numerical values of the SM-like Higgs
couplings and breaks electroweak gauge invariance.  The modified
dimension-4 coupling terms obviously affect all loop-induced Higgs
couplings. In addition, the Lagrangian includes new higher-dimensional
operators coupling the Higgs to photons and gluons.  They arise from
potential new particles in the loop and are normalized to their
Standard Model values $F_G$ and $F_A$. In the limit of heavy top
masses these normalization constants read $F_G^{(\infty)} \to
\alpha_s/(12 \pi)$.  We refrain from including a more complete set of
operators, to be consistent with existing analyses.  The form of
Eq.\eqref{eq:lag_delta} can be trivially mapped onto an effective
Lagrangian with a non-linear representation of the Higgs and Goldstone
fields, resulting in a broken $SU(2)$ doublet
structure~\cite{Buchalla:2012qq,Buchalla:2013eza,Brivio:2013pma,Brivio:2016fzo}.

Including invisible Higgs decays in terms of an effective Lagrangian
would force us to either define a new particle with unknown quantum
numbers, or to re-scale the decay $H \to 4 \nu$ to ridiculous
branching ratios. Instead, we include invisible Higgs decays in terms
of the corresponding partial width or, equivalently, the invisible
branching ratio. The total Higgs width is consistently constructed out
of all observed partial width, with an assumed scaling of the second
and third generation of Yukawa couplings, as described in
Ref.~\cite{Corbett:2015ksa}.

In principle it would be possible to include the Higgs self-coupling
in the global, non-linear Higgs analysis. However, we know that
di-Higgs measurements will not improve any of the parameters given in
Eq.\eqref{eq:lag_delta} and that the Higgs self-coupling does not
affect single-Higgs production in a relevant way.  Because the
extration of the Higgs self-coupling crucially depends on kinematic
distributions~\cite{Kling:2016lay,Goncalves:2018yva} we postpone this
aspect to Sec.~\ref{sec:gauge}, where we also include a full set of
di-boson and single Higgs distributions.\medskip

The global Higgs analysis in terms of Eq.\eqref{eq:lag_delta} only
describes total cross sections in the Higgs sector. We therefore just
re-scale the number of signal and background events in the 8~TeV
analysis~\cite{Corbett:2015ksa} to 27~TeV, assuming two
experiments. This affects all statistical uncertainties, as well as
the systematics, which we assume to be related to measurements in
control regions. For all measurements we assume the SM predictions,
which means that our best-fit points will always be the SM values.
For the invisible Higgs searches we use an in-house extrapolation of
the WBF analysis from Ref.~\cite{Biekotter:2017gyu} to 27~TeV.  The
current theory uncertainties of all measurements are listed in
Tab.~\ref{tab:uncert}, including uncertainties on the parton
distributions.  For these, we simply assume that dedicated fits will
determine the PDFs at a 27~TeV collider will full luminosity at the
same level as they are determined for the LHC now.  To illustrate the
point that precision predictions and PDF extraction are crucial we
will show results with the current theory uncertainties as well as an
assumed improvement of theory and systematics by a factor two.\medskip

In Fig.~\ref{fig:delta} we show the expected precision of the SM-like
Higgs coupling measurements for a 27~TeV LHC upgrade. Asymmetric
uncertainty bands arise because of correlations, but also reflect
numerical uncertainties.  Different colors correspond to assumed
integrated luminosities of $1.5~\iab$ and $15~\iab$.  For all coupling
deviations, with the exception of $\Delta_b$, we observe an
improvement with increased luminosity. 
However, this improvement is much smaller than the rough factor three which 
one could expect from a scaling of the limits with the square-root of the 
luminosity, indicating that the limits are systematics and theory limited.
Ratios of couplings, like
$\Delta_{W/Z}$, see little improvement from an increased luminosity.
To confirm the domination by systematics and
theory uncertainties, we also compare today's theory and systematic
uncertainties with an improvement to half the current uncertainties
indicated by full and shaded bands. For example for $\Delta_b$ and
$\Delta_g$, as well as for the coupling ratios we indeed see a
significant improvement.  Altogether, the typical precision in
measuring Higgs couplings can reach 3\% to 5\% at a 27~TeV hadron
collider with a realistic improvement of the systematic and theory
uncertainties. The ratio of the $W$ and $Z$ couplings to the Higgs,
for example, will benefit from correlated uncertainties and will
therefore be measured a factor two more precisely than the individual
couplings.  Invisible Higgs decays will be constrained at the
branching ratio level of 1\% to 2\%.  Compared to the high-luminosity
LHC predictions in Ref.~\cite{Klute:2013cx} the 27~TeV with its
projected final luminosity will double the precision on many of the
Higgs coupling modifications.

\section{Higgs-gauge analysis} 
\label{sec:gauge}

An effective Lagrangian is defined by its particle content and its
symmetries, with the dimensionality of the individual operators, or
inverse powers of a large matching scale as the expansion parameter.
Truncated to dimension six it has the
form~\cite{Hagiwara:1993ck,Grzadkowski:2010es}
\begin{align}
\lag = \sum_x \frac{f_x}{\Lambda^2} \; \ope_x \; ,
\label{eq:def_f}
\end{align}
where $\Lambda \gg v$ is the scale of the assumed UV-complete model.
It can be extended to the full Standard Model particle content,
defining the Standard Model Effective Field
Theory~\cite{Brivio:2017vri}.  The minimum independent set of
dimension-6 operators with the SM particle content, compatible with
the SM gauge symmetries, and compatible with baryon number
conservation contains 59 operators~\cite{Grzadkowski:2010es}.  We
impose $C$ and $P$ invariance~\cite{Corbett:2012dm} on the operator
set of Ref.~\cite{Hagiwara:1993ck}, use the equations of motion
including all necessary fermionic operators to avoid blind directions
from electroweak precision data, and neglect all operators that will
not be constrained by LHC Higgs measurements. For the Higgs couplings
to fermions we assume a Yukawa coupling structure, noting that current
LHC analyses are unlikely to test the Lorentz structure of a possible
deviation from the Standard Model. This gives us the Lagrangian
\begin{align}
\lag_\text{eff} 
= & - \frac{\alpha_s }{8 \pi} \frac{f_{GG}}{\Lambda^2} \ope_{GG}  
    + \frac{f_{BB}}{\Lambda^2} \ope_{BB} 
    + \frac{f_{WW}}{\Lambda^2} \ope_{WW} 
    + \frac{f_B}{\Lambda^2} \ope_B 
    + \frac{f_W}{\Lambda^2} \ope_W 
    + \frac{f_{WWW}}{\Lambda^2} \ope_{WWW} \notag \\ 
  &  + \frac{f_{\phi 2}}{\Lambda^2} \ope_{\phi 2} 
    + \frac{f_{\phi 3}}{\Lambda^2} \ope_{\phi 3} 
    + \frac{f_\tau m_\tau}{v \Lambda^2} \ope_{e\phi,33} 
    + \frac{f_b m_b}{v \Lambda^2} \ope_{d\phi,33} 
    + \frac{f_t m_t}{v \Lambda^2} \ope_{u\phi,33} \notag \\
  & + \text{invisible decays} \; ,
\label{eq:ourleff}
\end{align}
with the operators defined as in Ref.\cite{Butter:2016cvz},
\begin{alignat}{9}
\ope_{GG} &= \phi^\dagger \phi \; G_{\mu\nu}^a G^{a\mu\nu}  \quad 
&\ope_{BB} &= \phi^{\dagger} \hat{B}_{\mu \nu} \hat{B}^{\mu \nu} \phi 
&\ope_{WW} &= \phi^{\dagger} \hat{W}_{\mu \nu} \hat{W}^{\mu \nu} \phi  \quad 
\notag \\
\ope_B &=  (D_{\mu} \phi)^{\dagger}  \hat{B}^{\mu \nu}  (D_{\nu} \phi) 
&\ope_W &= (D_{\mu} \phi)^{\dagger}  \hat{W}^{\mu \nu}  (D_{\nu} \phi)
&\ope_{WWW} &= \tr \left( \hat{W}_{\mu \nu} \hat{W}^{\nu \rho} 
\hat{W}_\rho^\mu \right)  
\notag \\
\ope_{\phi 2} &= \frac{1}{2} \partial^\mu ( \phi^\dagger \phi )
                            \partial_\mu ( \phi^\dagger \phi ) \quad
& \ope_{\phi 3} &= - (\phi^\dagger \phi)^3/3
\label{eq:operators}   \\
\ope_{e\phi,33} &= \phi^\dagger\phi \; \bar L_3 \phi e_{R,3}  \qquad 
&\ope_{d\phi,33} &= \phi^\dagger\phi \; \bar Q_3 \phi d_{R,3}  \qqquad  
&\ope_{u\phi,33} &= \phi^\dagger\phi  \; \bar Q_3 \tilde \phi u_{R,3} \notag \; .
\end{alignat}
%
The pure gauge operator $\ope_{WWW}$ is needed to fully describe
anomalous triple gauge couplings in a gauge-invariant framework.  Any
LHC analysis also needs to include an anomalous triple gluon
interaction, but this operator is constrained by multi-jet production
at 13~TeV much more strongly than any Higgs analysis will
achieve~\cite{Krauss:2016ely}, and we assume that this pattern will be
the same at a 27~TeV collider.  Recent
studies~\cite{Franceschini:2017xkh,Banerjee:2018bio,Biekotter:2018rhp}
have shown the relevance of fermionic operators through their induced
$f \bar f V$, $f \bar f VV$ and $f \bar f VH$ couplings, which
require a combination of the Higgs--gauge analysis with electoweak
precision data. Such an analysis is beyond the scope of this
projection, but eventually the fermionic operators should be part of a
global SMEFT fit.\medskip

\begin{table}[b!]
\centering
\begin{tabular}{lrrr}
\toprule
channel  & observable & \# bins & range [GeV] \\
\midrule
$WW \rightarrow (\ell \nu)(\ell \nu)$          & $m_{\ell\ell'}$ & $10$ & $0-4500$\\
$WW \rightarrow (\ell \nu)(\ell \nu)$          & $p_T^{\ell_1}$ & $8$ & $0-1750$\\
$WZ \rightarrow (\ell \nu)(\ell \ell)$          & $m_T^{WZ}$ & $11$ & $0-5000$\\
$WZ \rightarrow (\ell \nu)(\ell \ell)$          & $p_T^{\ell\ell}$ ($p_T^Z$) & $9$ & $0-2400$\\
WBF, $H\rightarrow \gamma\gamma$          & $p_T^{\ell_1}$ & $9$ & $0-2400$\\
$VH \rightarrow (0\ell) (b \bar{b})$ & $p_T^V$ & $7$ & $150-750$\\
$VH \rightarrow (1\ell) (b \bar{b})$ & $p_T^V$ & $7$ & $150-750$\\
$VH \rightarrow (2\ell) (b \bar{b})$ & $p_T^V$ & $7$ & $150-750$\\
$HH \rightarrow (b \bar{b}) (\gamma \gamma)$, $2j$ & $m_{HH}$ & $9$ & $200-1000$ \\
$HH \rightarrow (b \bar{b}) (\gamma \gamma)$, $3j$ & $m_{HH}$ & $9$ & $200-1000$ \\
\bottomrule
\end{tabular}
\caption{Distributions included in the analysis. The number of bins
includes an overflow bin for all channels.}
\label{tab:distr}
\end{table}

Because the effective Lagrangian of Eq.\eqref{eq:ourleff} includes new
Lorentz structures, especially valuable information comes from
kinematic distributions probing interactions with a large momentum
flow.  We therefore include four single Higgs and four di-boson
distributions in our analysis~\cite{Butter:2016cvz}. For the 8~TeV
analysis they are validated with existing
data~\cite{Butter:2016cvz,Aad:2013wqa,Chatrchyan:2013lba}. We use
\textsc{MadGraph}5-2.3.2.2~\cite{Alwall:2014hca},
\textsc{Pythia}6-2.4.5~\cite{Sjostrand:2006za}, and
\textsc{Delphes}3.1.2~\cite{deFavereau:2013fsa} for two ATLAS/CMS-like
experiments.  For distributions, where different cuts define different
phase space regions, we only use the high-momentum regime. Finally, we
include Higgs pair production $pp \to HH \to b\bar{b}\gamma\gamma$
including kinematic information in terms of $m_{HH}$, as pioneered in
Ref.~\cite{Baur:2003gp}, now accounting for two different jet
multiplicities~\cite{Goncalves:2018yva}.  All distributions entering
our analysis are summarized in Tab.~\ref{tab:distr}, the details of
the Higgs pair production process will be discussed in
Sec.~\ref{sec:self}.

Comparing the reach of the main kinematic distributions in
Tab.~\ref{tab:distr} we see that the $VV$ channels probe a much larger
momentum flow than the $VH$ channels. This can be traced to the larger
signal rates, namely $\sigma_{WZ} = 61.1$~pb vs $\sigma_{WH} = 2.8$~pb
at 27~TeV and to leading order~\cite{Alwall:2014hca}, combined with
higher tails through the momentum-dependent $WWZ$ coupling. For example
comparing $WZ$ production with $WH$ production, the reach in $p_{T}^V$
is defined by the highest bins with a sizeable number of signal
events. Specifically, we ignore phase space regions with fewer than
three signal events for an integrated luminosity of $15~\iab$.\medskip

\begin{figure}[t]
\centering
\includegraphics[width=0.80\textwidth]{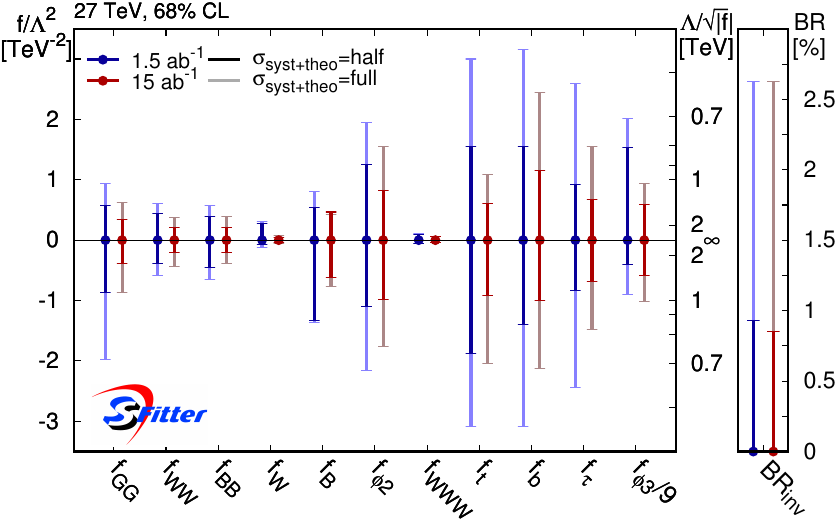}
\caption{Result from the global Higgs analysis in terms of dimension-6
  operators. All limits are shown as profiled over all other Wilson
  coefficients.}
\label{fig:d6}
\end{figure}

In Fig.~\ref{fig:d6} we show the results of the global Higgs analysis
in terms of dimension-6 SMEFT operators, including the quadratic terms
of the EFT expansion. The right axis indicates the new-physics scale
$\Lambda$ assuming reasonably strongly interacting new physics $f_x
=1$. The 27~TeV analysis is typically sensitive to new-physics scales
well above 1~TeV.  This number should be compared to the range of
distributions given in Tab.~\ref{tab:distr}, indicating that our Higgs
analysis does not have any serious EFT validity issues provided we do
not see a pole in the diboson channels.  The asymmetric error bands
for some of the Wilson coefficients will be discussed in the next
section.

The balance of statistical, systematic, and theory uncertainties in
the SMEFT analysis is significantly different from the non-linear
coupling modifiers shown in Fig.~\ref{fig:delta}. Effective operators
benefit from an increased statistics, because larger luminosity
extends the reach of kinematic distributions, which in their tails are
always statistically limited. In contrast, the Yukawa couplings
$f_{b,\tau}$, which do not change the Lorentz structure, are mostly
limited by the assumed systematic and theory uncertainties.
Consequently, the reach for operators which modify the Lorentz
structure of some Higgs interaction exceeds the reach for the
Yukawa-like operators or the reach for the operator $\ope_{\phi 2}$,
which introduces a wave function renormalization for the Higgs field
and only changes the kinematics of Higgs pair production. For the
former the kinematic distributions drive the limits towards
$\Lambda/\sqrt{f} \approx 3$~TeV and beyond, for high luminosity and
improved systematics and theory uncertainties. This aspect is where we
also expect significant improvements from a dedicated 27~TeV study
developing analysis ideas not realized at 8~TeV.\medskip

The asymmetric limits on $f_B$ give us some insight into the structure
of the effective theory. This operator is largely constrained through
$VH$ production at high momentum transfer, specifically the $p_T^V$
distributions from Tab.~\ref{tab:distr}. In its highest available bins
we probe sizeable ratios $p_T/\Lambda$, but with sizeable statistical
uncertainties. If we include the dimension-6 squared terms a second
solution predicting the same event count within the statistical
uncertainties appears for $f_B/\Lambda > 0$. For this second solution
the squared term compensates a small destructive interference with the
SM contribution. Because the precise position of this secondary
solution differs for different values of $p_{T,V}$, it induces a
slightly asymmetric measurement of $f_B/\Lambda$. Note that a visible
dimension-6-squared term in a specific observable does by no means
signal the breakdown of the effective
Lagrangian~\cite{Contino:2016jqw}. The validity of an effective field
theory representing classes of underlying UV-complete models can only
be judged once we identify on-shell contributions of the new
particles~\cite{Biekotter:2016ecg}. Second, truncating the expansion
of our observables after the linear term in $f/\Lambda^2$ would lead
to a symmetric and more narrow likelihood distribution and
underestimate of the errors.  In general, we do not include
uncertainties on the EFT framework in our global analysis, as we
consider them to be uncertainties on the matching and interpretation
of our results in terms of a UV complete
model~\cite{Brehmer:2015rna,Helset:2018dht}.

\section{Higgs self-interaction}
\label{sec:self}

\begin{figure}[b!]
\centering
\includegraphics[width=0.55\textwidth]{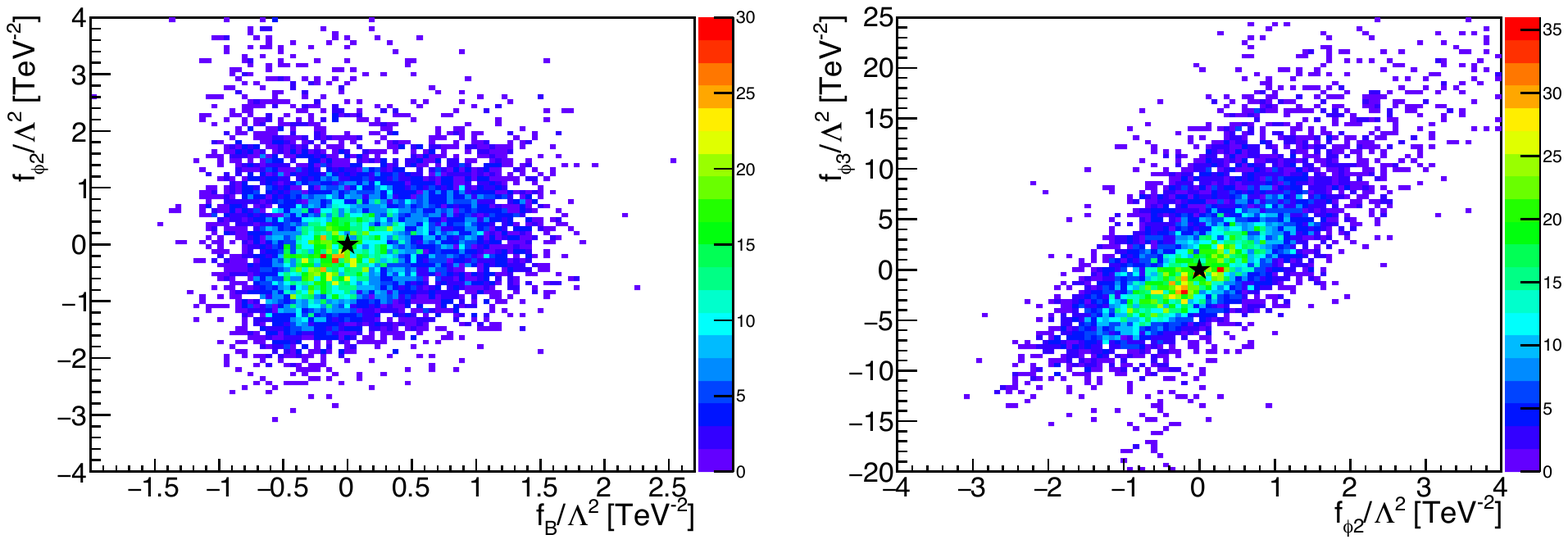}
\caption{Correlations between the leading operators describing Higgs
  pair production.}
\label{fig:corr}
\end{figure}

An enhanced Higgs self-coupling as a simple modification of the SM
Higgs potential is especially interesting for example in relation to
vacuum stability and
baryogenesis~\cite{Grojean:2004xa,Reichert:2017puo}.  Including it in
our global Higgs analysis is a significant improvement as compared to
the Run~I legacy analysis~\cite{Butter:2016cvz}. It is made possible
by the fact that a 27~TeV collider with a large integrated luminosity
will allow for a dedicated measurement of the Higgs self-coupling.
The self-coupling with its unique relation to the Higgs potential is
not yet included in most global analyses of SM-like Higgs couplings
because of the modest reach of the LHC.  However, for a 27~TeV
collider with an integrated luminosity of $15~\iab$ we quote the
expected reach~\cite{Goncalves:2018yva}
\begin{align}
\frac{\lambda_{3H}}{\lambda_{3H}^\text{(SM)}}
=\begin{cases} 
1 \pm 15\% &  68\%~\text{C.L.} \\
1 \pm 30\% &  95\%~\text{C.L.} 
\end{cases}
\label{eq:reach_lam}
\end{align}
We can translate this range into the conventions of
Eq.\eqref{eq:ourleff} if we assume that the underlying new physics
does not generate any other dimension-6 operator. In that case we
assume
\begin{align}
V = \mu^2 \; \frac{(v+H)^2}{2} 
  + \lambda \; \frac{(v+H)^4}{4} 
  + \frac{f_{\phi 3}}{3 \Lambda^2} \; \frac{(v+H)^6}{8} \; ,
\end{align}
and find for the reach of the dedicated self-coupling
analysis~\cite{Plehn:2009nd}
\begin{align}
\lambda_{3H} 
= \lambda_{3H}^\text{(SM)}
  \left( 1 + \frac{2 v^2}{3 m_H^2} \; \frac{f_{\phi 3} v^2}{\Lambda^2} \right) 
\qquad \text{and} \qquad
\left| \frac{\Lambda}{\sqrt{f_{\phi 3}}} \right| \gtrsim
\begin{cases}
1~\tev &  68\%~\text{C.L.} \\
700~\gev &  95\%~\text{C.L.} 
\end{cases}
\label{eq:reach_d6}
\end{align}
While we are free to define a modified Higgs potential as our physics
hypothesis~\cite{Reichert:2017puo}, this setup is in direct violation
of the effective Lagrangian approach. Here all operators consistent
with the symmetry assumptions have to be included. Consequently,
$\ope_{\phi 2}$ also affects the Higgs pair production process with a
momentum-dependent self-coupling~\cite{Plehn:2009nd,Barger:2003rs}.\medskip

The full kinematic information from Higgs pair production encoded in
the $m_{HH}$ distribution allows us to separate the effects of
$\ope_{\phi 2}$ and $\ope_{\phi
  3}$~\cite{Kling:2016lay,Goncalves:2018yva}. In addition, we can use
single Higgs production to constrain $\ope_{GG}$ and
$\ope_{u\phi,33}$, as seen in Sec.~\ref{sec:gauge}.  All corresponding
error bands have to be propagated into Higgs pair production, and
given the size of the uncertainties we can safely assume that in the
presence of $\ope_{\phi 3}$ Higgs pair production will hardly help
with any of the operators already constrained by single Higgs
production.  Following Refs.~\cite{Bizon:2016wgr,Maltoni:2017ims} and
especially Ref.~\cite{DiVita:2017eyz} we also neglect the loop effects
of $\ope_{\phi 3}$ on single Higgs production, because they will
hardly affect a global Higgs analysis.  Finally, looking at different
uncertainties it is also obvious that a study of the $m_{HH}$
distribution will be statistically limited even at the 27~TeV
collider, in contrast to the typical total rate measurements discussed
before.

\begin{figure}[t]
\centering
\includegraphics[width=0.72\textwidth]{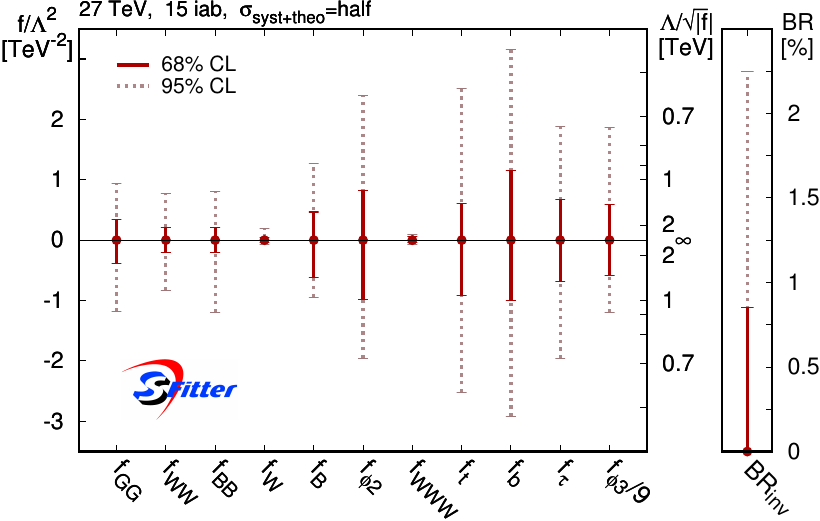}
\caption{Result from the global Higgs analysis in terms of dimension-6
  operators, complementing the high-luminosity and improved-error
  scenario of Fig.~\ref{fig:d6} with results for 95\% C.L. or two
  standard deviations.}
\label{fig:95cl}
\end{figure}  

In Fig.~\ref{fig:corr} we illustrate the correlation of $\ope_{\phi
  2}$ and $\ope_{\phi 3}$ through the full expression for a modified
self-coupling~\cite{Plehn:2009nd}.  We also know that positive and
negative deviations of the Higgs self-coupling affect different phase
space regions~\cite{Kling:2016lay}: while a reduced value of the
self-coupling can be tested around the threshold $m_{HH} \approx 2
m_H$, an increase in the self-coupling requires us to look for large
values of $m_{HH}$ and to deal with large effects from $\ope_{\phi
  2}$. All of this leads to an asymmetric uncertainty band on
$\ope_{\phi 3}$ especially once we allow for an agreement of the
SM-like measurements with the SM predictions at two sigma and
integrate the asymmetric tails further. From Fig.~\ref{fig:95cl} we
read off the limits
\begin{alignat}{9}
\frac{\Lambda}{\sqrt{|f_{\phi 3}|}} &> 430~\gev
&\qquad \text{68\% C.L.} \notag \\
\frac{\Lambda}{\sqrt{|f_{\phi 3}|}} &> 245~\gev
\quad (f_{\phi 3} >0)
\quad \text{and} \quad 
\frac{\Lambda}{\sqrt{|f_{\phi 3}|}} > 300~\gev
\quad (f_{\phi 3} <0) 
&\qquad \text{95\% C.L.} 
\label{eq:reach_d6_3}
\end{alignat}
These limits are diluted from the one-parameter analysis quoted in
Eq.\eqref{eq:reach_d6}, largely because of the combination with
$\ope_{\phi 2}$. As a matter of fact, we can directly compare the
effects from $\ope_{\phi 2}$ and $\ope_{\phi 3}$ for similar values of
$f/\Lambda^2$ as a function of the momentum flowing through the
triple-Higgs vertex or $m_{HH}$. In that case we find that the
momentum dependence in $\ope_{\phi 2}$ matches the effects from
$\ope_{\phi 3}$ for $m_{HH} \gtrsim 1$~TeV, with either relative
sign. This additional source of a modified self-coupling characterized
by the interplay between $\ope_{\phi 2}$ and $\ope_{\phi 3}$ is not
accounted for in the usual Higgs pair analyses.

\section{Outlook} 

Following the established Run~I legacy
results~\cite{Corbett:2015ksa,Butter:2016cvz} we estimate the reach of
a 27~TeV hadron collider in a global analysis of the Higgs-gauge
sector. We include not only invisible Higgs decays, but also Higgs
pair production, sensitive to the Higgs self-coupling.

First, we interpret the extrapolated measurement in terms of modified
SM-like Higgs couplings, motivated by an effective theory with
non-linearly realized electroweak symmetry breaking. We find that a
27~TeV hadron collider will be sensitive to $3~...~5\%$ deviations
from the SM coupling values. Systematics and theory uncertainties
rapidly limit the reach for modified Higgs couplings beyond
attobarn-level integrated luminosities.

Using a gauge-invariant effective theory in terms of the Higgs doublet
allows us to include di-boson rates and kinematic distributions in the
global Higgs analysis.  Invisible branching ratios can be extracted
from a global Higgs analysis to better than one per-cent.  For the
gauge-invariant interpretation framework we also include a modified
Higgs potential at dimension six. The additional Wilson coefficient
can be constrained by a kinematic analysis of Higgs pair
production~\cite{Goncalves:2018yva} if we control the correlation
with the operator $\ope_{\phi 2}$.

We find a TeV-scale reach for new physics given order-one
Wilson coefficients for most dimension-6 operators. Those operators
which change the Lorentz structure of Higgs couplings can be strongly
constrained beyond the 3~TeV level.  
Comparing those numbers with the current reach of the 
LHC Run~II~\cite{Ellis:2018gqa,Alves:2018nof,Biekotter:2018rhp}, we find that the reach of 
a  27~TeV hadron collider could increase the bounds on the new physics
scale by more than $50\%$.
The global analysis obviously
reduces the reach for the Higgs self-coupling modification compared to
one-parameter analysis, but still indicates that a 27~TeV hadron
collider will for the first time deliver a meaningful measurement of
this fundamental physics parameter. 

\bigskip
\newpage
{\begin{center} \textbf{Acknowledgments} \end{center}
\nopagebreak
First and foremost we are grateful to Michael Rauch, not only because
of his significant contribution in the early phase of this project,
but also because of his long-term leading role in all SFitter
analyses. If our field decides to let technically cutting-edge people
like him go, we should not be surprised that large parts of BSM theory
are losing contact to LHC data and are becoming increasingly
irrelevant to LHC physics as a whole.

We would like to thank Jennie Thompson for providing the numbers for
the invisible Higgs analysis. AB is funded by the DFG through the
Graduiertenkolleg Particle physics beyond the Standard Model (GRK
1940) and the IMPRS-PTFS.  The authors acknowledge support by the
state of Baden-W\"urttemberg through bwHPC and the German Research
Foundation (DFG) through grant no INST 39/963-1 FUGG (bwForCluster
NEMO).  MT is supported in part by the JSPS Grant-in-Aid for
Scientific Research Numbers~16H03991, 16H02176, 17H05399, 18K03611,
and the World Premier International Research Center Initiative, MEXT,
Japan.}

\bibliographystyle{mystyle}
\bibliography{references}

\begin{thebibliography}{10}
\expandafter\ifx\csname url\endcsname\relax
  \def\url#1{\texttt{#1}}\fi
\expandafter\ifx\csname doi\endcsname\relax
  \def\doi#1{\burlalt{doi:#1}{http://dx.doi.org/#1}}\fi
\expandafter\ifx\csname urlprefix\endcsname\relax\def\urlprefix{URL }\fi
\expandafter\ifx\csname href\endcsname\relax
  \def\href#1#2{#2}\fi
\expandafter\ifx\csname burlalt\endcsname\relax
  \def\burlalt#1#2{\href{#2}{#1}}\fi

\bibitem{Dawson:2018dcd}
S.~Dawson, C.~Englert, and T.~Plehn,
  \burlalt{1808.01324}{http://arxiv.org/abs/1808.01324}.

\bibitem{Leung:1984ni}
C.~N. Leung, S.~T. Love, and S.~Rao, Z. Phys., C31 (1986) 433.

\bibitem{Buchmuller:1985jz}
W.~Buchmuller and D.~Wyler, Nucl. Phys., B268 (1986) 621--653.

\bibitem{DeRujula:1991ufe}
A.~De~Rujula, M.~B. Gavela, P.~Hernandez, and E.~Masso, Nucl. Phys., B384
  (1992) 3--58.

\bibitem{Hagiwara:1993qt}
K.~Hagiwara, R.~Szalapski, and D.~Zeppenfeld, Phys. Lett., B318 (1993)
  155--162, \burlalt{hep-ph/9308347}{http://arxiv.org/abs/hep-ph/9308347}.

\bibitem{Hagiwara:1993ck}
K.~Hagiwara, S.~Ishihara, R.~Szalapski, and D.~Zeppenfeld, Phys. Rev., D48
  (1993) 2182--2203.

\bibitem{Hagiwara:1995vp}
K.~Hagiwara, S.~Matsumoto, and R.~Szalapski, Phys. Lett., B357 (1995) 411--418,
  \burlalt{hep-ph/9505322}{http://arxiv.org/abs/hep-ph/9505322}.

\bibitem{GonzalezGarcia:1999fq}
M.~C. Gonzalez-Garcia, Int. J. Mod. Phys., A14 (1999) 3121--3156,
  \burlalt{hep-ph/9902321}{http://arxiv.org/abs/hep-ph/9902321}.

\bibitem{Grzadkowski:2010es}
B.~Grzadkowski, M.~Iskrzynski, M.~Misiak, and J.~Rosiek, JHEP, 10 (2010) 085,
  \burlalt{1008.4884}{http://arxiv.org/abs/1008.4884}.

\bibitem{Passarino:2012cb}
G.~Passarino, Nucl. Phys., B868 (2013) 416--458,
  \burlalt{1209.5538}{http://arxiv.org/abs/1209.5538}.

\bibitem{Brivio:2017vri}
I.~Brivio and M.~Trott, \burlalt{1706.08945}{http://arxiv.org/abs/1706.08945}.

\bibitem{Corbett:2015ksa}
T.~Corbett, O.~J.~P. Eboli, D.~Goncalves, J.~Gonzalez-Fraile, T.~Plehn, and
  M.~Rauch, JHEP, 08 (2015) 156,
  \burlalt{1505.05516}{http://arxiv.org/abs/1505.05516}.

\bibitem{Corbett:2012ja}
T.~Corbett, O.~J.~P. Eboli, J.~Gonzalez-Fraile, and M.~C. Gonzalez-Garcia,
  Phys. Rev., D87 (2013) 015022,
  \burlalt{1211.4580}{http://arxiv.org/abs/1211.4580}.

\bibitem{Banerjee:2013apa}
S.~Banerjee, S.~Mukhopadhyay, and B.~Mukhopadhyaya, Phys. Rev., D89(5) (2014)
  053010, \burlalt{1308.4860}{http://arxiv.org/abs/1308.4860}.

\bibitem{Ellis:2014jta}
J.~Ellis, V.~Sanz, and T.~You, JHEP, 03 (2015) 157,
  \burlalt{1410.7703}{http://arxiv.org/abs/1410.7703}.

\bibitem{Ellis:2018gqa}
J.~Ellis, C.~W. Murphy, V.~Sanz, and T.~You, JHEP, 06 (2018) 146,
  \burlalt{1803.03252}{http://arxiv.org/abs/1803.03252}.

\bibitem{Brivio:2013pma}
I.~Brivio, T.~Corbett, O.~J.~P. Éboli, M.~B. Gavela, J.~Gonzalez-Fraile, M.~C.
  Gonzalez-Garcia, L.~Merlo, and S.~Rigolin, JHEP, 03 (2014) 024,
  \burlalt{1311.1823}{http://arxiv.org/abs/1311.1823}.

\bibitem{Falkowski:2015jaa}
A.~Falkowski, M.~Gonzalez-Alonso, A.~Greljo, and D.~Marzocca, Phys. Rev. Lett.,
  116(1) (2016) 011801, \burlalt{1508.00581}{http://arxiv.org/abs/1508.00581}.

\bibitem{Falkowski:2016cxu}
A.~Falkowski, M.~Gonzalez-Alonso, A.~Greljo, D.~Marzocca, and M.~Son, JHEP, 02
  (2017) 115, \burlalt{1609.06312}{http://arxiv.org/abs/1609.06312}.

\bibitem{Berthier:2016tkq}
L.~Berthier, M.~Bjørn, and M.~Trott, JHEP, 09 (2016) 157,
  \burlalt{1606.06693}{http://arxiv.org/abs/1606.06693}.

\bibitem{Liu:2018pkg}
D.~Liu and L.-T. Wang, \burlalt{1804.08688}{http://arxiv.org/abs/1804.08688}.

\bibitem{Baglio:2017bfe}
J.~Baglio, S.~Dawson, and I.~M. Lewis, Phys. Rev., D96(7) (2017) 073003,
  \burlalt{1708.03332}{http://arxiv.org/abs/1708.03332}.

\bibitem{Franceschini:2017xkh}
R.~Franceschini, G.~Panico, A.~Pomarol, F.~Riva, and A.~Wulzer, JHEP, 02 (2018)
  111, \burlalt{1712.01310}{http://arxiv.org/abs/1712.01310}.

\bibitem{Kling:2016lay}
F.~Kling, T.~Plehn, and P.~Schichtel, Phys. Rev., D95(3) (2017) 035026,
  \burlalt{1607.07441}{http://arxiv.org/abs/1607.07441}.

\bibitem{Goncalves:2018yva}
D.~Gonçalves, T.~Han, F.~Kling, T.~Plehn, and M.~Takeuchi, Phys. Rev., D97(11)
  (2018) 113004, \burlalt{1802.04319}{http://arxiv.org/abs/1802.04319}.

\bibitem{Kilian:2018bhs}
W.~Kilian, S.~Sun, Q.-S. Yan, X.~Zhao, and Z.~Zhao,
  \burlalt{1808.05534}{http://arxiv.org/abs/1808.05534}.

\bibitem{Bizon:2018syu}
W.~Bizoń, U.~Haisch, and L.~Rottoli,
  \burlalt{1810.04665}{http://arxiv.org/abs/1810.04665}.

\bibitem{Homiller:2018dgu}
S.~Homiller and P.~Meade,
  \burlalt{1811.02572}{http://arxiv.org/abs/1811.02572}.

\bibitem{Grojean:2004xa}
C.~Grojean, G.~Servant, and J.~D. Wells, Phys. Rev., D71 (2005) 036001,
  \burlalt{hep-ph/0407019}{http://arxiv.org/abs/hep-ph/0407019}.

\bibitem{Reichert:2017puo}
M.~Reichert, A.~Eichhorn, H.~Gies, J.~M. Pawlowski, T.~Plehn, and M.~M.
  Scherer, Phys. Rev., D97(7) (2018) 075008,
  \burlalt{1711.00019}{http://arxiv.org/abs/1711.00019}.

\bibitem{Biekotter:2017gyu}
A.~Biekötter, F.~Keilbach, R.~Moutafis, T.~Plehn, and J.~Thompson, SciPost
  Phys., 4(6) (2018) 035,
  \burlalt{1712.03973}{http://arxiv.org/abs/1712.03973}.

\bibitem{Butter:2016cvz}
A.~Butter, O.~J.~P. Éboli, J.~Gonzalez-Fraile, M.~C. Gonzalez-Garcia,
  T.~Plehn, and M.~Rauch, JHEP, 07 (2016) 152,
  \burlalt{1604.03105}{http://arxiv.org/abs/1604.03105}.

\bibitem{Lafaye:2009vr}
R.~Lafaye, T.~Plehn, M.~Rauch, D.~Zerwas, and M.~Duhrssen, JHEP, 08 (2009) 009,
  \burlalt{0904.3866}{http://arxiv.org/abs/0904.3866}.

\bibitem{deBlas:2016ojx}
J.~de~Blas, M.~Ciuchini, E.~Franco, S.~Mishima, M.~Pierini, L.~Reina, and
  L.~Silvestrini, JHEP, 12 (2016) 135,
  \burlalt{1608.01509}{http://arxiv.org/abs/1608.01509}.

\bibitem{Alves:2018nof}
A.~Alves, N.~Rosa-Agostinho, O.~J.~P. Éboli, and M.~C. Gonzalez-Garcia, Phys.
  Rev., D98(1) (2018) 013006,
  \burlalt{1805.11108}{http://arxiv.org/abs/1805.11108}.

\bibitem{Biekotter:2018rhp}
A.~Biekötter, T.~Corbett, and T.~Plehn,
  \burlalt{1812.07587}{http://arxiv.org/abs/1812.07587}.

\bibitem{Englert:2015hrx}
C.~Englert, R.~Kogler, H.~Schulz, and M.~Spannowsky, Eur. Phys. J., C76(7)
  (2016) 393, \burlalt{1511.05170}{http://arxiv.org/abs/1511.05170}.

\bibitem{deFlorian:2016spz}
D.~de~Florian et~al., \burlalt{1610.07922}{http://arxiv.org/abs/1610.07922}.

\bibitem{Zeppenfeld:2000td}
D.~Zeppenfeld, R.~Kinnunen, A.~Nikitenko, and E.~Richter-Was, Phys. Rev., D62
  (2000) 013009, \burlalt{hep-ph/0002036}{http://arxiv.org/abs/hep-ph/0002036}.

\bibitem{Duhrssen:2004cv}
M.~Duhrssen, S.~Heinemeyer, H.~Logan, D.~Rainwater, G.~Weiglein, and
  D.~Zeppenfeld, Phys. Rev., D70 (2004) 113009,
  \burlalt{hep-ph/0406323}{http://arxiv.org/abs/hep-ph/0406323}.

\bibitem{Buchalla:2012qq}
G.~Buchalla and O.~Cata, JHEP, 07 (2012) 101,
  \burlalt{1203.6510}{http://arxiv.org/abs/1203.6510}.

\bibitem{Buchalla:2013eza}
G.~Buchalla, O.~Catá, and C.~Krause, Phys. Lett., B731 (2014) 80--86,
  \burlalt{1312.5624}{http://arxiv.org/abs/1312.5624}.

\bibitem{Brivio:2016fzo}
I.~Brivio, J.~Gonzalez-Fraile, M.~C. Gonzalez-Garcia, and L.~Merlo, Eur. Phys.
  J., C76(7) (2016) 416, \burlalt{1604.06801}{http://arxiv.org/abs/1604.06801}.

\bibitem{Klute:2013cx}
M.~Klute, R.~Lafaye, T.~Plehn, M.~Rauch, and D.~Zerwas, EPL, 101(5) (2013)
  51001, \burlalt{1301.1322}{http://arxiv.org/abs/1301.1322}.

\bibitem{Corbett:2012dm}
T.~Corbett, O.~J.~P. Eboli, J.~Gonzalez-Fraile, and M.~C. Gonzalez-Garcia,
  Phys. Rev., D86 (2012) 075013,
  \burlalt{1207.1344}{http://arxiv.org/abs/1207.1344}.

\bibitem{Krauss:2016ely}
F.~Krauss, S.~Kuttimalai, and T.~Plehn, Phys. Rev., D95(3) (2017) 035024,
  \burlalt{1611.00767}{http://arxiv.org/abs/1611.00767}.

\bibitem{Banerjee:2018bio}
S.~Banerjee, C.~Englert, R.~S. Gupta, and M.~Spannowsky, Phys. Rev., D98(9)
  (2018) 095012, \burlalt{1807.01796}{http://arxiv.org/abs/1807.01796}.

\bibitem{Aad:2013wqa}
G.~Aad et~al., Phys. Lett., B726 (2013) 88--119,
  \burlalt{1307.1427}{http://arxiv.org/abs/1307.1427}.
\newblock [Erratum: Phys. Lett.B734,406(2014)].

\bibitem{Chatrchyan:2013lba}
S.~Chatrchyan et~al., JHEP, 06 (2013) 081,
  \burlalt{1303.4571}{http://arxiv.org/abs/1303.4571}.

\bibitem{Alwall:2014hca}
J.~Alwall, R.~Frederix, S.~Frixione, V.~Hirschi, F.~Maltoni, O.~Mattelaer,
  H.~S. Shao, T.~Stelzer, P.~Torrielli, and M.~Zaro, JHEP, 07 (2014) 079,
  \burlalt{1405.0301}{http://arxiv.org/abs/1405.0301}.

\bibitem{Sjostrand:2006za}
T.~Sjostrand, S.~Mrenna, and P.~Z. Skands, JHEP, 05 (2006) 026,
  \burlalt{hep-ph/0603175}{http://arxiv.org/abs/hep-ph/0603175}.

\bibitem{deFavereau:2013fsa}
J.~de~Favereau, C.~Delaere, P.~Demin, A.~Giammanco, V.~Lemaître, A.~Mertens,
  and M.~Selvaggi, JHEP, 02 (2014) 057,
  \burlalt{1307.6346}{http://arxiv.org/abs/1307.6346}.

\bibitem{Baur:2003gp}
U.~Baur, T.~Plehn, and D.~L. Rainwater, Phys. Rev., D69 (2004) 053004,
  \burlalt{hep-ph/0310056}{http://arxiv.org/abs/hep-ph/0310056}.

\bibitem{Contino:2016jqw}
R.~Contino, A.~Falkowski, F.~Goertz, C.~Grojean, and F.~Riva, JHEP, 07 (2016)
  144, \burlalt{1604.06444}{http://arxiv.org/abs/1604.06444}.

\bibitem{Biekotter:2016ecg}
A.~Biekötter, J.~Brehmer, and T.~Plehn, Phys. Rev., D94(5) (2016) 055032,
  \burlalt{1602.05202}{http://arxiv.org/abs/1602.05202}.

\bibitem{Brehmer:2015rna}
J.~Brehmer, A.~Freitas, D.~Lopez-Val, and T.~Plehn, Phys. Rev., D93(7) (2016)
  075014, \burlalt{1510.03443}{http://arxiv.org/abs/1510.03443}.

\bibitem{Helset:2018dht}
A.~Helset and M.~Trott, \burlalt{1812.02991}{http://arxiv.org/abs/1812.02991}.

\bibitem{Plehn:2009nd}
T.~Plehn, Lect. Notes Phys., 844 (2012) 1--193,
  \burlalt{0910.4182}{http://arxiv.org/abs/0910.4182}.

\bibitem{Barger:2003rs}
V.~Barger, T.~Han, P.~Langacker, B.~McElrath, and P.~Zerwas, Phys. Rev., D67
  (2003) 115001, \burlalt{hep-ph/0301097}{http://arxiv.org/abs/hep-ph/0301097}.

\bibitem{Bizon:2016wgr}
W.~Bizon, M.~Gorbahn, U.~Haisch, and G.~Zanderighi, JHEP, 07 (2017) 083,
  \burlalt{1610.05771}{http://arxiv.org/abs/1610.05771}.

\bibitem{Maltoni:2017ims}
F.~Maltoni, D.~Pagani, A.~Shivaji, and X.~Zhao, Eur. Phys. J., C77(12) (2017)
  887, \burlalt{1709.08649}{http://arxiv.org/abs/1709.08649}.

\bibitem{DiVita:2017eyz}
S.~Di~Vita, C.~Grojean, G.~Panico, M.~Riembau, and T.~Vantalon, JHEP, 09 (2017)
  069, \burlalt{1704.01953}{http://arxiv.org/abs/1704.01953}.

\end{thebibliography}

\end{document}